\documentstyle[psfig]{elsart}

\newcommand{\epr}{e-print}

\begin{document} 
\begin{frontmatter}

\title{Lorentz  Invariance  Violation  and  the Observed  Spectrum  of
Ultrahigh Energy Cosmic Rays}

\author{S.T. Scully}  \address{Department of Physics  and Astronomy \\
James  Madison   University,  Harrisonburg,  VA   22807}  \author{F.W.
Stecker} \address{Astrophysics  Science Division\\ NASA  Goddard Space
Flight Center, Greenbelt, MD 20771}

\begin{abstract} 

There  has  been  much  interest  in possible  violations  of  Lorentz
invariance,  particularly motivated by  quantum gravity  theories.  It
has been suggested that a small amount of Lorentz invariance violation
(LIV)  could  turn off  photomeson  interactions  of ultrahigh  energy
cosmic rays  (UHECRs) with photons of the  cosmic background radiation
and thereby  eliminate the resulting sharp steepening  in the spectrum
of the  highest energy  CRs predicted by  Greisen Zatsepin  and Kuzmin
(GZK). Recent measurements of the UHECR spectrum reported by the HiRes
and Auger  collaborations, however, indicate  the presence of  the GZK
effect.   We present  the results  of  a detailed  calculation of  the
modification of the  UHECR spectrum caused by LIV  using the formalism
of  Coleman and  Glashow.   We  then compare  these  results with  the
experimental  UHECR data  from Auger  and HiRes. Based on these data, 
we find a best fit amount of  LIV
of $4.5^{+1.5}_{-4.5} \times 10^{-23}$,consistent with an upper limit 
of $6 \times 10^{-23}$. This possible 
amount of LIV can  lead to  a recovery of the cosmic ray spectrum at 
higher energies than presently observed. Such an LIV recovery 
effect can be tested observationally using future detectors.

\end{abstract} 


\end{frontmatter}

\section{Introduction}

Because  of their  extreme energy  and isotropic  distribution,  it is
believed that UHECRs are extragalactic in origin.  After the discovery
of  the cosmic  background  radiation (CBR),  Greisen \cite{gr66}  and
Zatsepin   and  Kuzmin   \cite{za66}  pointed   out   that  photomeson
interactions  should deplete  the flux  of cosmic  rays  with energies
above $\sim$ 50 EeV.  One of  us \cite{st68}, using data on the energy
dependence  of the photomeson  production cross  section, then  made a
quantitative  calculation of  this  ``GZK effect''  deriving the  mean
photomeson energy loss attenuation length for protons as a function of
proton energy.  These results indicated that the attenuation length of
a proton  with an  energy greater than  100 EeV  is less than  100 Mpc,
which is  much less  than the visible  radius of the  universe.  Thus,
what is  sometimes referred  to as  the GZK ``cutoff''  is not  a true
cutoff,  but a  suppression of  the ultrahigh  energy cosmic  ray flux
arising from a limitation of the proton propagation length through the
cosmic background radiation owing to energy losses.

From time  to time there  have been reports  in the literature  of the
detection  of giant  air shower  events from  primaries  with energies
above the GZK suppression  energy (trans-GZK events) ({\it e.g.}, Refs.
~\cite{li63} -- \cite{ta03}). Such  events have stimulated 
suggestions  that a
violation  of Lorentz  invariance  or a  modification  of the  Lorentz
transformation  relations  at ultrahigh  energies  could  result in  a
nullification  of  the   GZK  effect \cite{sa72},\cite{ki72}.   Most
significantly,  the  AGASA group  reported  11  events  above the  GZK
suppression  energy   \cite{ta03},  increasing  the   interest  in  the
possibility of such new physics \cite{st03a}. See Ref.\cite{bi08} for 
a recent review of this topic. 

However, a reanalysis of the  AGASA data (unpublished) has resulted in
cutting their originally reported  number of trans-GZK events by half.
More importantly, the HiRes \cite{ab08} and Auger groups \cite{abr08},
with larger  exposures, have very  recently claimed to have  found a
GZK  suppression  effect.  Motivated  by  these  new  results, we  have
undertaken new  detailed calculations  of the effect  of a  very small
amount of Lorentz invariance violation (LIV) on the spectrum of UHECRs
at Earth. We present our results  here and compare them with the HiRes
and Auger data separately.

\section{Violating Lorentz Invariance}  

With  the idea of  spontaneous symmetry  breaking in  particle physics
came  the suggestion  that  Lorentz invariance  (LI)  might be  weakly
broken at  high energies. Some of  the more recent  motivation for LIV
has been in relating it  to possible Planck scale phenomena that could
lead to astrophysically observable consequences~\cite{ac98}.  The idea
that Planck scale physics may lead  to a natural abrogation of the GZK
effect has  been of  particular interest, since  this would lead  to a
direct observational test.  Significant  fluxes of UHECRs at trans-GZK
energies,  could  be  the  result  of  a  very  small  amount  of  LIV
~\cite{co99}  --  ~\cite{al03}.   Such  a test  would  have  important
implications  for  some  quantum  gravity and  large  extra  dimension
models, since those models may predict very small amount of LIV.

Although no  true quantum theory of  gravity exists, it  is natural to
tie LIV  to various  quantum gravity models.   A few examples  of such
work  can be  found in  Refs.~\cite{al00} --  ~\cite{sm08}.   For more
references, we refer to  an excellent review by Mattingly \cite{ma05}.
A data table of constraints on LIV and CPT violation parameters within
the  framework of the  ``Standard Model  Extension'' model~\cite{ck98}
has recently been given by Kostelecky and Russell ~\cite{ko08}.

In this paper, we reinvestigate the observational implications of the
possible effect of a very small amount of LIV, {\it viz.}, that cosmic
rays could indeed reach us after originating at distances greater than
100  Mpc  without  undergoing  large  energy  losses  from  photomeson
interactions.  We  considered this topic  before in a  more simplistic
manner~\cite{st05}  when there  was  a clear  discrepancy between  the
AGASA group  data~\cite{ta03} and the earlier HiRes  data. However, as
discussed  above,  the observational  situation  has  changed and  now
requires a  more detailed approach. We therefore  undertook a detailed
calculation of  the modification of  the UHECR spectrum caused  by LIV
using  the  formalism  of  Coleman  and Glashow  and  the  kinematical
approach  originally  given by  Alfaro  and  Palma~\cite{al03} in  the
context  of the  Loop  Quantum Gravity  model~\cite{ab92},\cite{sm05}.
(See also Ref.~\cite{al05}.)
Then, by comparing our results with the observational UHECR data we can
place a quantitative limit on the  amount of LIV. 
We also  discuss how  a small amount  of LIV  that is
consistent with the observational data  can still lead to a recovery of
the cosmic ray flux at higher energies than presently observed.

\section{LIV Framework}

Coleman and Glashow have proposed a simple formulation for breaking LI
by   a  small   first  order   perturbation  in   the   free  particle
Lagrangian~\cite{co99}.   This  formalism has  the  advantages of  (1)
simplicity,  (2) preserving  the  $SU(3) \otimes  SU(2) \otimes  U(1)$
standard model of strong  and electroweak interactions, (3) having the
perturbative term  in the Lagrangian to consist of operators of mass 
dimension $4$ that
thus  preserves  power   counting  renormalizability,  and  (4)  being
rotationally invariant  in a preferred frame  that can be  taken to be
the  rest  frame of  the  2.7  K  cosmic background  radiation.   This
formalism  has  proven  useful  in exploring  astrophysical  data  for
testing LIV ~\cite{co99},\cite{ja04},\cite{sg01}.

To accomplish this,  Coleman and Glashow start with  the free particle
Lagrangian

\begin{equation}
{\cal L} = \partial_{\mu} \Psi ^ * {\bf Z} \partial^{\mu}\Psi - \Psi ^ *
{\bf M}^2\Psi 
\end{equation}

where $\Psi$ is a column vector of $n$ fields with U(1) invariance and the
positive  Hermitian  matrices  ${\bf   Z}$  and  {\bf  M}$^2$  can  be
transformed  so that  ${\bf Z}$  is the  identity and  {\bf  M}$^2$ is
diagonalized  to  produce the  standard  theory  of  $n$ decoupled  free
fields.

They then add a leading order perturbative, Lorentz violating term 
constructed from only spatial derivatives so that

\begin{equation}
{{\cal L} \rightarrow {\cal L} + 
\partial_i\Psi {\bf \epsilon} \partial^i\Psi} ,  
\end{equation}

where {\bf $\epsilon$} is a dimensionless Hermitian matrix that 
commutes with {\bf M}$^2$
so that the fields remain separable and the resulting single particle
energy-momentum eigenstates go from eigenstates of {\bf M}$^2$ at low energy 
to eigenstates of {\bf $\epsilon$} at high energies.

The Lorentz violating perturbative term shifts the poles of the propagators, 
resulting in free particle dispersion relations of the form

\begin{equation}
E^2 ~ = ~ \vec{p} \ ^{2} + m^{2} + \epsilon \vec{p} \ ^2.
\label{dispersion}
\end{equation}

These can be put in the standard form for the dispersion relations

\begin{equation}
E^2 ~ =~  \vec{p \ }{^2}c_{MAV}^2 + m^{2} c_{MAV}^4,
\end{equation}

by shifting the renormalized mass by the small amount 
$m \rightarrow m/(1+\epsilon)$ and shifting the velocity from c (=1)
by the amount $c_{MAV} = \sqrt{(1 +  \epsilon)} \simeq 1 + \epsilon/2$.

Since the group velocity is given by

\begin{equation}
{{\partial E}\over{\partial |\vec{p}|}} = {{|\vec{p}|} \over {\sqrt {|\vec{p}|^2 + m^2 c_{MAV} ^2}}} c_{MAV} ~~
 \rightarrow ~ ~ c_{MAV}~~ {\rm as} ~~ |\vec{p}|~~ \rightarrow~ \infty ,
\label{MAV}
\end{equation}

Coleman and
Glashow thus identify $c_{MAV}$ as the maximum attainable velocity of the
free particle.
 
Using  this formalism, it becomes apparent that, in principle,
different particles  can have different  maximum attainable velocities
(MAVs) resulting from the individually distinguishable eigenstates of 
the {\bf $\epsilon$ } matrix. These various MAVs can all be different  from $c$
as well as different from each other.  Hereafter, we denote the MAV  of a
particle  of type $i$  by $c_{i}$  and the  difference 

\begin{equation}
c_{i}  - c_{j} ~=~ {{\epsilon_{i}-\epsilon_{j}}\over{2}}~ \equiv ~ \delta_{ij}
\label{deltadef}
\end{equation}

There are other popular formalisms that are inspired by quantum 
gravity models or by speculations on the nature of space-time at 
the Planck scale, $1/M_{Pl} \simeq 1.5 \times
10^{-35}$ m, where $M_{Pl} = 1/\sqrt G \simeq 1.2 \times 10^{19}$ GeV. 
Such formalisms, in the context of effective field theory, can be 
expressed by postulating Lagrangians containing operators of dimension$\ge$ 5 
with suppression factors as multiples of $M_{Pl}~$\cite{ja04},\cite{mp03}. 
This leads to dispersion relations having a series of smaller and
smaller terms proportional to 
$p^{n+2}/M_{Pl}^n \simeq E^{n+2}/M_{Pl}^n$, with $n \ge 1$. However,
in relating LIV to the observational data on UHECRs, we find it useful
to use the simpler formalism of Coleman and Glashow. Given the
limited energy range of the UHECR data relevant to the GZK effect,
this formalism can later be related to possible Planck scale 
phenomena and quantum gravity models of various sorts.

We now consider the  photomeson 
production process near threshold where single pion production dominates,

\begin{equation}
p + \gamma \rightarrow N + \pi.
\end{equation}

Using the normal Lorentz invariant kinematics, the energy threshold for
photomeson interactions of UHECR protons of initial laboratory energy $E$ with 
low energy photons of the CBR  with laboratory
energy $\omega$ is determined by the relativistic invariance of the 
square of the total four-momentum of the proton-photon system. This relation,
together with the threshold inelasticity relation $E_{\pi} = [m/(M + m)]E$ for
single pion production, yields the threshold conditions for head on collisions
in the laboratory frame. In terms of the pion energy for single pion 
production at threshold

\begin{equation}
4\omega E_{\pi} ~ = ~ {{m^2(2M + m)}  \over {M + m}} ,
\label{pion}
\end{equation}

where M is the rest mass of the proton and 
m is the rest mass of the pion~\cite{st68}.

If LI is broken so that $c_\pi~ >~ c_p$, 
it follows from equations (\ref{dispersion}), (\ref{deltadef})
and (\ref{pion}) that the threshold energy for  
photomeson production is altered 
because the square of the four-momentum is shifted from its LI form 
so that the threshold condition becomes

\begin{equation}
4\omega E_{\pi} ~  = ~ {{m^2(2M + m)} \over {M + m}} + 2 \delta_{\pi p} E_{\pi}^2
\label{LIVpi}
\end{equation}

Equation (\ref{LIVpi}) is a quadratic equation with real roots only
under the condition 

\begin{equation}
\delta_{\pi p} \le {{2\omega^2(M + m)} \over {m^2(2M + m)}} \simeq 
  \omega^2/m^2.
\label{root}
\end{equation}

Defining $\omega_0 \equiv kT = 2.35 \times 10^{-4}$ eV, equation (\ref{root})
can be rewritten

\begin{equation}
\delta_{\pi p} \le 3.23 \times 10^{-24} (\omega/\omega_0)^2.
\label{CG}
\end{equation}

If LIV occurs and $\delta_{\pi p} > 0$, photomeson production can only
take place for interactions of  CBR photons with energies large enough
to satisfy equation (\ref{CG}). Single photon photomeson production 
takes is dominated by the $\Delta$ resonance and takes place close 
to the interaction threshold. This fact, together with equation
(\ref{LIVpi}) implies that under some conditions photomeson interactions
leading to GZK  suppression can occur for ``lower  energy'' UHE protons
interacting with relatively higher energy CBR photons on the Wien tail
of  the  Planck spectrum,  but  such  interactions  for higher  energy
protons,  which  would normally  interact  with  photons having  smaller
values  of  $\omega$,  will  be  forbidden. Thus,  the  observed  UHECR
spectrum may  exhibit the characteristics  of GZK suppression  near the
normal GZK threshold, but the UHECR spectrum can ``recover'' at higher
energies  owing to  the  possibility that  photomeson interactions  at
higher  proton  energies  may   be  forbidden.

\section{Kinematics}

We now consider a  
detailed quantitative treatment of this possibility, {\it viz., 
GZK coexisting with LIV}. 
We first give the kinematical relations needed  to perform our
calculations  in  the  presence   of  a  small  violation  of  Lorentz
invariance.  We will  denote quantities in the proton  rest frame by a
prime and quantities in the cms system of the proton-photon collision
by  an  asterisk.   Quantities   in  the  laboratory  frame  are  left
unprimed. Following equations (\ref{dispersion}) and (\ref{deltadef}), 
we denote

\begin{equation}
E^2=p^2+2\delta _a p^2 +{m_a}^2
\end{equation}

where $\delta _a$  is the difference between the  MAV for the particle
{\it  a} and  the speed  of light in the low momentum limit ($c = 1$).

The cms energy of particle $a$ is then given by

\begin{equation}
\sqrt{s_a} = \sqrt{E^2-p^2} = \sqrt{2\delta_a p^2 + m_a^2}
\label{restmass}
\end{equation}

where, of course, we must have the condition  $s_a \ge 0$.
It is important to note that, owing to LIV, in the cms where p = 0
the particle will not generally be at rest because

\begin{equation}
v = {{\partial E} \over {\partial p}} \neq {{p}\over {E}}.
\end{equation}

We follow Ref. \cite{al03} in defining the square of the {\it total} rest 
energy in the cms by

\begin{equation}
s \equiv  E_{tot}^2 - p_{tot}^2 
\end{equation}

Then denoting $s_p$ to be the square of the initial proton energy in the system
where the initial proton momentum is zero, it follows that

\begin{equation}
s = 2\sqrt{s_p}\epsilon + s_p
\end{equation}

where we now define $\epsilon$ as the energy of the photon in this system.

Now let us  obtain the expression for the  modified inelasticity, $K$,
for the  photopion producing reaction $p+\gamma \rightarrow  N + \pi$.
Since the inelasticity is defined  by the fraction of the total energy
carried away  by the pion,  we can relate  the energy of  the emerging
proton  and  pion  to  the  total  energy  in  the  laboratory  system
(essentially the initial energy of the proton) by
\begin{eqnarray}
E_{\pi}   &  =   &  K_{\theta}   E_p  \nonumber   \\  E_{N}   &   =  &
(1-K_{\theta})E_{p}
\end{eqnarray}
where $K_{\theta}$ is  the inelasticity for a given  $\theta$ which is
the angle between the momentum vectors of the photon and the proton in
the laboratory  system.  In  order to solve  for the  inelasticity, we
calculate the  cms energy of the  nucleon in two  different ways.  On
one  hand, we  can use  the Lorentz  transformation of  the laboratory
nucleon energy to relate it to the cms energy:
\begin{eqnarray}
E_{N} & = & \gamma^*({E_N}^*+\beta^*{p_N}^*\cos \theta) \nonumber \\ &
      = & \gamma^*({E_N}^*+\beta^*\sqrt{{E_N}^*-s_N(E_N)}\cos \theta) ,
      \label{ener1} 
\end{eqnarray}
where the Lorentz  factor for the cms frame is the  ratio of the total
laboratory energy $E_{p}+\omega \approx E_{p}$ to the total cms energy
which is given by equation (\ref{restmass})  and where
we now define $\omega$ to be  the observed energy of the CBR photon in
the laboratory system..

On the  other hand, we can derive  the cms energy of  the nucleon from
the threshold conditions by replacing the masses of the particles with
their rest energies as  prescribed by equation (\ref{restmass}).  This
yields the relationship
\begin{equation}
2\sqrt{s}{E_N}^*=s + s_N - s_{\pi}
\label{ener2} 
\end{equation}
where  the  quantities $s_N$  and  $s_{\pi}$  can  be determined  from
equation (\ref{restmass}) and are given by
\begin{eqnarray}
s_{\pi} &  = & \delta _\pi  (K_{\theta} E_{p_i})^2 +{m_\pi}^2\nonumber
\\ s_N & = & \delta _N [(1-K_{\theta})E_{p_i}]^2 + {m_p}^2.
\end{eqnarray}

Here we have replaced $p$ with  $E$ since we can exchange momentum for
energy, given the  high Lorentz factor.  We can  now combine equations
(\ref{ener1}) and (\ref{ener2}) to yield a transcendental equation for
$K_{\theta}$:
\begin{eqnarray}
\noindent
(1-K_{\theta})\sqrt           {s}            &           =           &
{{(s+s_N(K_{\theta})-s_\pi(K_{\theta}))}\over{2\sqrt{s}}}   \nonumber\\
&                                  +                                 &
\beta\sqrt{{(s+s_N(K_{\theta})-s_\pi(K_{\theta}))^2\over{4s}}-s_N}\cos
\theta .
\label{Kangle}
\end{eqnarray}
The total inelasticity,  $K$, will be an average  of $K_{\theta}$ with
respect to the angle between the proton and photon momenta, $\theta$:
\begin{equation}
K = \frac{1}{\pi }\int\limits_0^\pi {K_\theta d\theta }.
\label{Ktot}
\end{equation}

The primary  effect of LIV on  photopion production is  a reduction of
phase space allowed for the interaction.  This results from the limits
on  the  allowed range  of  interaction  angles  implied by  equations
(\ref{Kangle}) and  (\ref{Ktot}).  As the pion rest  energy grows, the
cosine term in equation (\ref{Kangle}) becomes larger.  For collisions
with  $\theta  <  \pi  /2$,  kinematically  allowed  solutions  become
severely restricted.   The modified  inelasticity that results  is the
key in  determining the  effects of LIV  on photopion  production. The
inelasticity rapidly drops for higher incident proton energies.

As shown in Ref.\cite{co99}, in order to modify the effect of 
photopion production on the UHECR spectrum above the GZK energy
we must have $\delta_{\pi} > \delta_p$. It is shown
in Figure 10 of Ref. ~\cite{jlm03} that for most of the allowed
parameter space near threshold  $\delta_{\pi}$ can be as much as
an order of magnitude greater than  $\delta_p$. Therefore, in this
paper we will assume that $\delta_{\pi} \gg \delta_p$ at or near
threshold. This assumption is also made in Ref. ~\cite{al03}.
We will thus take $\delta_{\pi p} \simeq \delta_{\pi} \equiv \delta$.
We have numerically determined that the dependence of our results on
the  $\delta_{\pi p}$ parameter
dominates over that on the $\delta_p$ parameter, as concluded in
Ref. \cite{co99}. The effect of taking a value of $\delta_p$ comparable to 
$\delta_{\pi}$ on the UHECR spectrum will be presented in a future paper. 

\begin{figure}
  \centerline{\psfig{figure=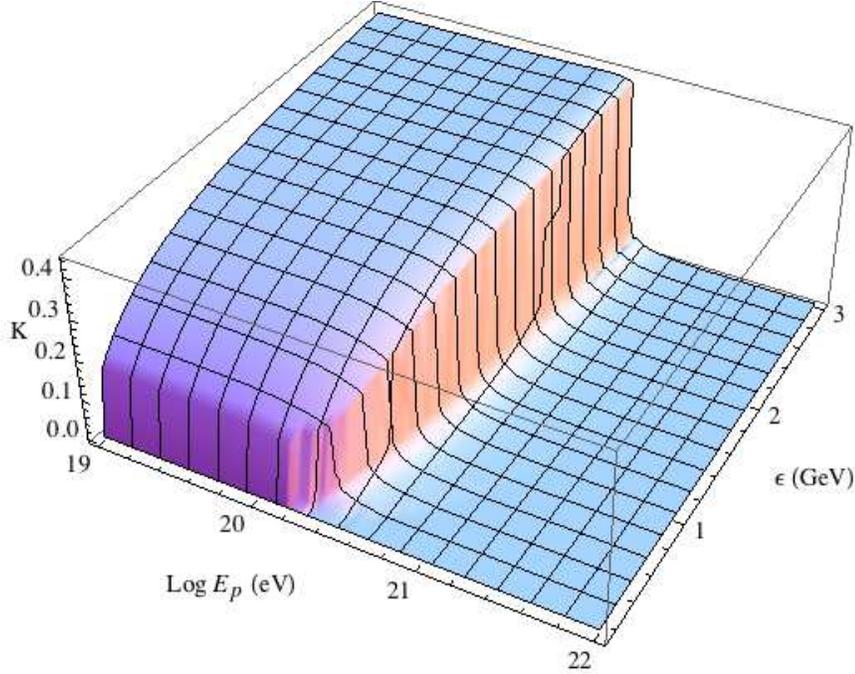,height=3.6in}}
  \caption{The  calculated  proton inelasticity  modified  by LIV  for
$\delta_{\pi p} =  3 \times 10^{-23}$ as  a function of CBR  photon energy and
proton energy.}
\label{inelasticity}
\end{figure}

Figure  \ref{inelasticity} shows  the  calculated proton  inelasticity
modified  by LIV for  a value  of $\delta_{\pi p}  = 3  \times 10^{-23}$  as a
function of  both CBR photon energy  and proton energy.   
Other choices for $\delta_{\pi p}$
yield similar  plots.  The principal  result of changing the  value of
$\delta_{\pi p}$  is  to  change  the  energy  at  which  LIV  effects  become
significant.  For a  choice of $\delta_{\pi p} = 3  \times 10^{-23}$, there is
no observable effect from LIV for $E_{p}$ less than $\sim 2 \times 10^{20}$ eV.
Above  this energy, the  inelasticity precipitously  drops as  the LIV
term in the pion rest energy approaches $m_{\pi}$.

With  this  modified inelasticity,  the  proton  energy  loss rate  by
photomeson production is given by

\begin{equation}
{{1}\over{E}}{{dE}\over{dt}}     =     -    {{\omega_{0}c}\over{2\pi^2
\gamma^2}\hbar^3c^3} \int\limits_\eta^\infty  d\epsilon ~ \epsilon ~
\sigma(\epsilon)                                            K(\epsilon)
\ln[1-e^{-\epsilon/2\gamma\omega_{0}}]\end{equation}

where $\eta$ is the photon threshold energy for the interaction in the
cms and $\sigma(\epsilon)$ is  the total $\gamma$-p cross section with
contributions from  direct pion production,  multipion production, and
the $\Delta$ resonance.

\begin{figure}
  \centerline{\psfig{figure=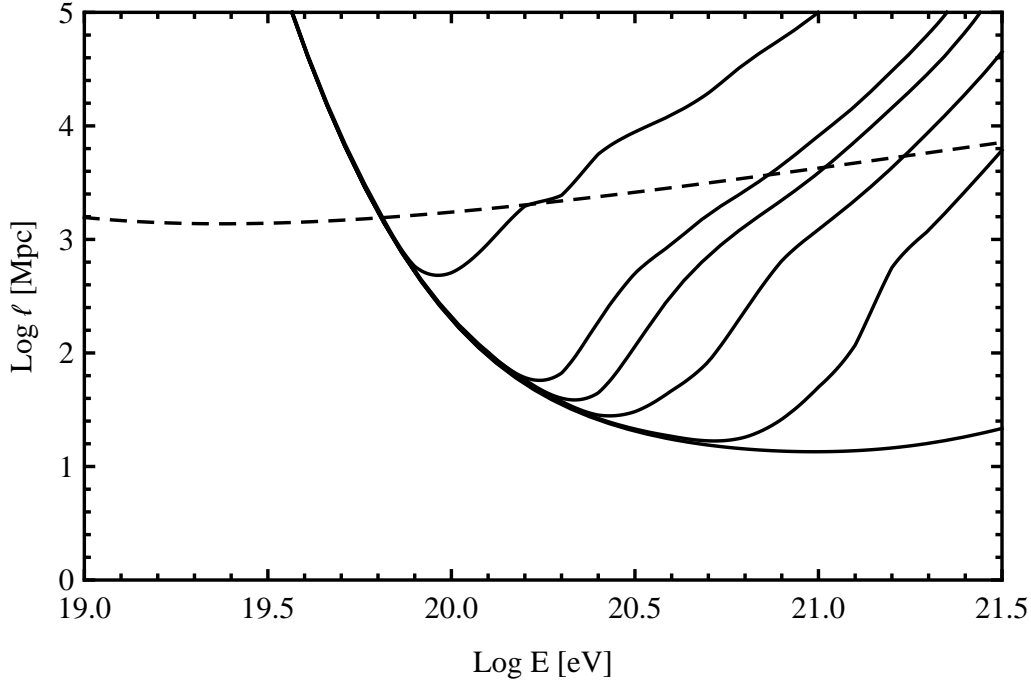,height=3.6in}}
  \caption{The  calculated  proton attenuation  lengths  as a  function
proton energy  modified by LIV  for various values of  $\delta_{\pi p}$
(solid lines), shown
with the attenuation length for pair production unmodified by LIV (dashed
lines). From top to bottom, the curves are for 
$\delta_{\pi p} = 1 \times 10^{-22}, 3 \times 10^{-23}
, 2 \times 10^{-23}, 1 \times 10^{-23}, 3 \times 10^{-24}, 
0$ (no Lorentz violation).}
\label{attn}
\end{figure}

The corresponding proton attenuation  length is given by $cE/(dE/dt)$.
This attenuation  length is plotted  in Figure \ref{attn}  for various
values of  $\delta_{\pi p}$ along with the  unmodified pair production
attenuation for comparison. We do not explore the effects of modifying
pair production through LIV in this paper.
 
\section{UHECR Spectra with LIV and Comparison with Present Observations}

We  will  start our  calculation  of  LIV  modified UHECR  spectra  by
assuming power-law  source spectra for  the UHECRs that are  chosen to
fit the UHECR data below 60 EeV.  We then consider the
propagation of high energy  protons, including energy losses resulting
from  cosmological redshifting,  pair production  and  pion production
through  interactions with  CBR  photons.  

We  shall  assume for  this
calculation  a flat $\Lambda$CDM  universe with  a Hubble  constant of
H$_0$ = 70 km s$^{-1}$ Mpc$^{-1}$, taking $\Omega_{\Lambda}$ = 0.7 and
$\Omega_{m}$  = 0.3. 
The energy  loss owing to redshifting for  a $\Lambda$CDM universe
is then given by
\begin{eqnarray}
-(\partial \log  E/\partial t)_{redshift} =  H_{0}\sqrt{\Omega_{m}(1+z)^3 +
  \Omega_{\Lambda}}.
  \label{eq3}
\end{eqnarray}
The attenuation length for protons against 
pair production  is given by

\begin{eqnarray}
-(\partial  \log  E/\partial  t)_{\gamma  p}  \equiv  r_{\gamma  p}  =
r_{\pi}(E) + r_{e^+e^-}(E),
  \label{eq4}
\end{eqnarray}

The attenuation lengths, $\ell  = cE/r(E)$, for protons against energy
loss by both pion production, with and without LIV, are shown together
with that  for pair production  in Figure \ref{attn}.  The  CBR photon
number  density increases  as $(1+z)^3$  and the  CBR  photon energies
increase  linearly with  $(1+z)$.  The  corresponding energy  loss for
protons at any redshift $z$ is thus given by

\begin{eqnarray}
r_{\gamma p}(E,z) = (1+z)^3 r[(1+z)E].
\label{eq5}
\end{eqnarray}
We  take   the  photomeson   loss  rate,  $r_{\pi}(E)$,   by  updating
~\cite{st68} using  the latest cross  sections listed in  the Particle
Data  Group ({\tt  http://pdg.lbl.gov}) and  in  Ref.~\cite{be07}.  We
take  the   pair-production  loss  rate,   $r_{e^+e^-}(E)$  from  Ref.
\cite{bl70}.

We  calculate the  initial energy,  $ E_i(z)$,  at which  a  proton is
created at a redshift $z$ whose observed energy today is $E$ following
the methods detailed in  Refs. \cite{be88} and \cite{sc02}. We neglect
the  effect of  possible small  intergalactic magnetic  fields  on the
paths  of these  ultrahigh energy  protons and  assume that  they will
propagate along  straight lines from  their source. The total  flux of
emitted particles from a volume element $dV=R^{3}(z)r^{2} dr d \Omega$
from redshift $z$  and distance $r$ with measured  energy $E$ is given
by
\begin{eqnarray} 
J(E)dE = {{q(E_{i},z)dE_{i}dV}\over{(1+z) 4 \pi R_{0}^{2} r^{2}}}.
\label{diffJ}
\end{eqnarray}
We assume that the average
UHECR volume emissivity is given by
$q(E_i,z) = K(z)E_i^{-\Gamma}$.

We will  assume a source  evolution $q(E_i, z)  \propto (1+z)^{\zeta}$
with $\zeta$ = 3.6, out to  a maximum redshift of 2.5. This assumption
corresponds to a  redshift evolution that is proportional  to the star
formation rate.  Our results on  LIV are insensitive to  the evolution
model assumed because evolution does not affect the shape of the UHECR
spectrum near the GZK cutoff energy ~\cite{st05,be88}. At higher energies 
where the attenuation length may again become large owing to an LIV
effect, we find the effect of evolution to be less than 10\% when
compared to the no-evolution case ($\zeta = 0$).

Since  $R_{0}  = (1+z)  R(z)$  and  $R(z)dr=cdt$,
by integrating equation (\ref{diffJ}), one obtains
\begin{eqnarray} 
J(E) = 
{{3cK(0)}\over{8\pi H_{0}}} E^{-\Gamma}\int_{0}^{z_{max}}{{(1+z)^{(\zeta-1)}}\over{\sqrt{\Omega_{m}(1+z)^3 +\Omega_{\Lambda}}}}
\left({E_i\over{E}}\right)^{-\Gamma}{{dE_i}\over{dE}}dz.
\end{eqnarray}
In this expression, K(0) is 
determined by fitting our final calculated spectrum to the observational 
UHECR data ~\cite{st05} assuming $\Gamma=2.55$, which  is
consistent with the Auger data below 100 EeV.

\begin{figure}
  \centerline{\psfig{figure=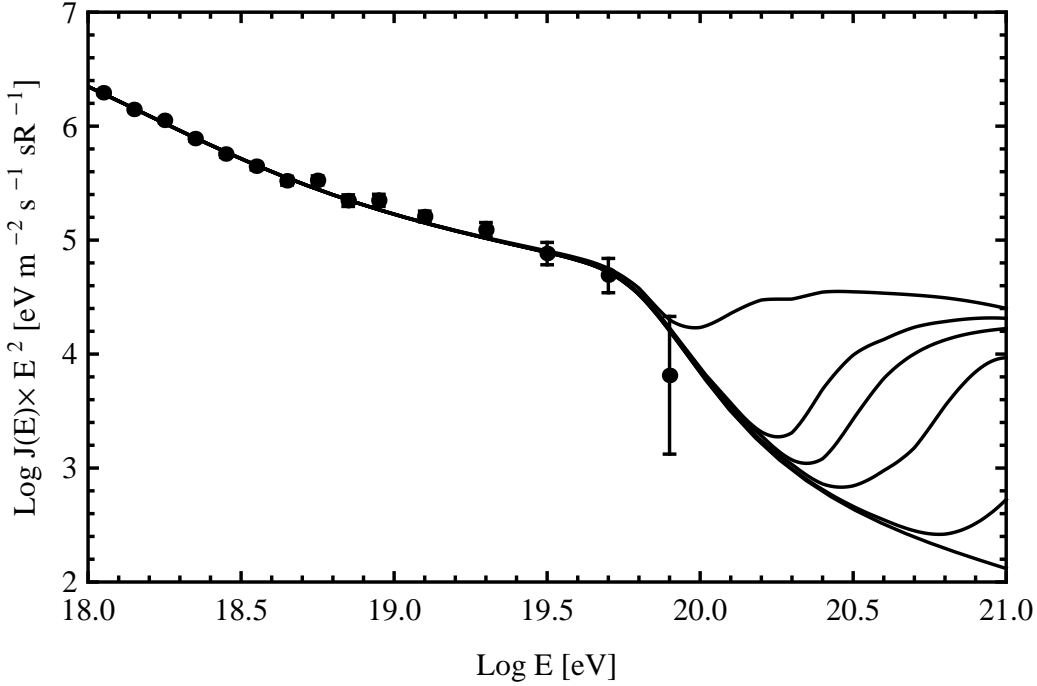,height=3.6in}}
  \caption{Comparison of the HiRes II data with calculated spectra for various 
values of $\delta_{\pi p}.$ From top to bottom, the curves give the predicted
spectra for 
$\delta_{\pi p} = 1 \times 10^{-22}, 3 \times 10^{-23}
, 2 \times 10^{-23}, 1 \times 10^{-23}, 3 \times 10^{-24}, 
0$ (no Lorentz violation).}
\label{HiRes}
\end{figure}

\begin{figure}
  \centerline{\psfig{figure=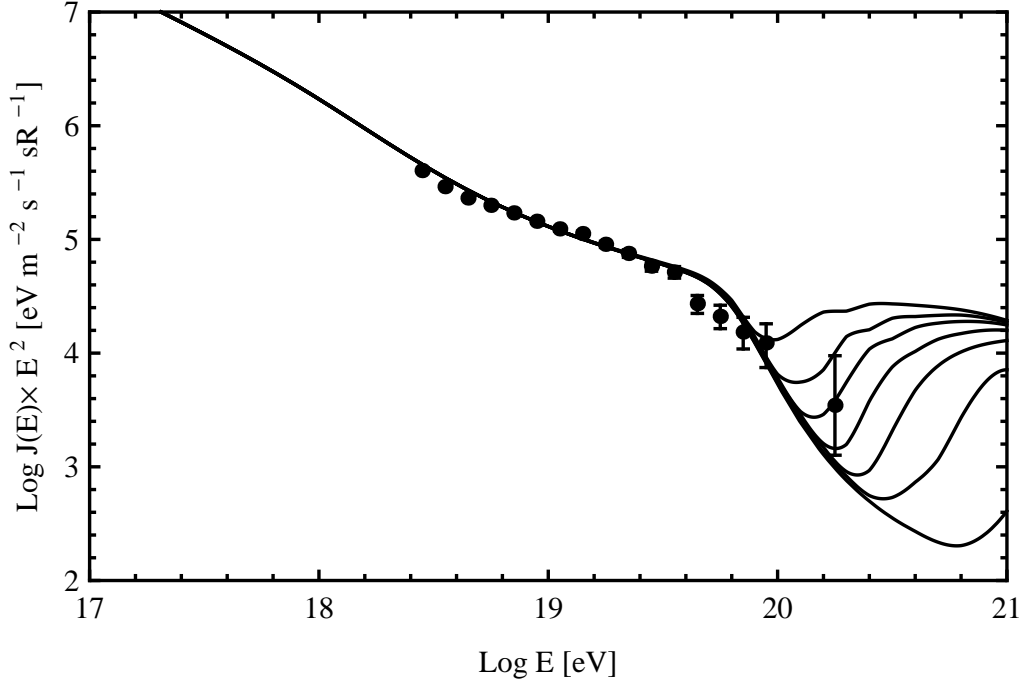,height=3.6in}}
  \caption{Comparison of the Auger data with calculated spectra for various 
values of $\delta_{\pi p}.$ From top to bottom, the curves give the predicted 
spectra for 
$\delta_{\pi p} = 1 \times 10^{-22}, 6 \times 10^{-23}, 4.5 \times 10^{-23}, 
3 \times 10^{-23} , 2 \times 10^{-23}, 1 \times 10^{-23}, 3 \times 10^{-24}, 
0$ (no Lorentz violation).}
\label{Auger}
\end{figure}

The results are shown in Figures \ref{HiRes} and \ref{Auger}.

\section{Discussion of Results}

It has been suggested that a  small  amount of  Lorentz
invariance   violation (LIV) could   turn  off   photomeson
interactions of ultrahigh energy cosmic rays (UHECRs) with photons of
the cosmic background radiation  and thereby eliminate the 
resulting sharp steepening in the spectrum of the highest energy CRs 
predicted by Greisen Zatsepin and Kuzmin (GZK). Recent measurements of the 
UHECR spectrum reported by the HiRes  \cite{ab08} and Auger \cite{abr08} 
collaborations, however, indicate the possible presence of a GZK effect. 

In order to determine the implications
for the search for Lorentz invariance violation at ultrahigh energies
from the analysis of the air shower events observed by HiRes and 
AGASA, we undertook a detailed analysis of the spectral features
produced by modifications of the kinematical relationships caused by
LIV at ultrahigh energies. In our analysis, we calculate modified 
UHECR spectra for various values of the Coleman-Glashow parameter,
$\delta_{\pi p}$, defined as the difference between the maximum attainable
velocities of the pion and the proton produced by LIV. We then 
compare our results with the experimental 
UHECR data and thereby place limits on the amount of LIV as defined by
the $\delta_{\pi p}$ parameter. 

Our results show that the amount of presently observed GZK suppression
in the UHECR data is consistent with the possible existence of a small
amount of LIV.  In order to  quantify this, we determined the value of
$\delta_{\pi p}$ that results in the smallest $\chi^2$ for the modeled
UHECR spectral fit using the  observational data above the GZK energy.
We find  this value to be  $4.5 \times 10^{-23}$.   We then determined
the range of acceptable values for $\delta_{\pi p}$.  This was done by
computing the probablity of getting a $\chi^2$ value at least as small
as  the   $\chi^2$  value  determined  from  the   fit.   We  rejected
$\delta_{\pi  p}$ values  outside of  the confidence  level associated
with 1$\sigma$.  We thus obtained  a best-fit range of
$\delta_{\pi p}$ = $4.5^{+1.5}_{-4.5} \times  10^{-23}$, corresponding
to an upper limit on $\delta_{\pi p}$ of $6 \times 10^{-23}$, as shown
in Figure \ref{Auger}.

The HiRes spectral data (see Figure \ref{HiRes}) do not
go to high enough energy to quantitatively constrain  
LIV. We also note that the Auger spectrum, being consistent with no 
obvious pair-production feature, does not constrain LIV for the 
pair-production interaction.
 
A small LIV effect can be distinguished from a higher energy component produced
by so-called top-down models because
the latter predict relatively large fluxes of UHE photons and neutrinos
as well as a significant diffuse GeV background flux that
could be searched for by the Fermi $\gamma$-ray space telescope. 
The Pierre Auger Observatory collaboration has provided observational
upper limits on the
UHE photon flux that have already disfavored top-down models \cite{au08}.
The upper limits form Auger indicate that UHE photons at best make up 
only a small percentage of the total UHE flux. This contradicts predictions
of top-down models that the flux of UHE photons should be larger than that
of UHE protons (See Ref.~\cite{st03a} for a review).   

As opposed to the predictions of the top-down models, the LIV effect
cuts off UHE pion production at the higher energies and consequent UHE 
neutrino and photon production from UHE pion decay. LIV would also not produce 
a GeV photon flux. 

It is also possible that the apparent modified GZK suppression 
in the data may be related to
an overdensity of nearby sources related to the local supergalactic
enhancement~\cite{st68}.\footnote{A correlation with nearby 
AGN has been hinted at in the Auger data~\cite{2ab08}. However, 
the HiRes group has found no significant correlation~\cite{abb08}.} 
More and better data will be required in order to resolve this question.
An LIV effect can be distinguished from a local source enhancement
by looking for UHECRs at energies above $\sim$200 EeV, as can be seen
from Figures \ref{HiRes} and \ref{Auger}. 
This is because the small amount of LIV that fits the observational UHECR
spectra can lead to a  recovery of the cosmic ray flux at higher
energies than  presently observed. Searching for such an effect
will require obtaining a data set containing
a much higher number of UHECR air shower events. 

In the  future, such  an increased number  of events may  be obtained.
The Auger collaboration has proposed to build an ``Auger North'' array
that would be seven times  larger than the present southern hemisphere
Auger  array  ({\tt  http://www.augernorth.org}).   Further  into  the
future,  space-based telescopes  designed  to look  downward at  large
areas of  the Earth's  atmosphere as a  sensitive detector  system for
giant air-showers caused by trans-GZK cosmic rays~\cite{st04}. We look
forward to these developments that may have important implications for
fundamental high energy physics.

\section*{Acknowledgment}

STS gratefully acknowledges
partial support from the Thomas F. \& Kate Miller Jeffress Memorial Trust 
grant no. J-805.

\end{document}